# INDUSTRIAL FABRICATION OF MEDIUM-BETA SCRF CAVITIES FOR A HIGH-INTENSITY PROTON LINAC

J. Kuzminski, General Atomics, San Diego, CA 92186, USA
K.C.D. Chan, R. Gentzlinger, LANL, Los Alamos, NM 87545 USA
P. Maccioni, CERCA, Romans, France

*Abstract*

During 1999, four 700-MHz, medium-beta ($\beta = 0.64$), superconducting radio frequency (SCRF) cavities for a high-intensity proton linac project at Los Alamos National Laboratory (LANL) were manufactured by industry. The SCRF cavities were designed by a LANL team in Los Alamos, New Mexico, USA, and manufactured at a CERCA plant in Romans, France. The cavities were made of 4-mm-thick, solid niobium sheets with a residual resistivity ratio (RRR) greater than 250. These niobium sheets were supplied by Wah Chang (USA), Heraeus AG (Germany), and Tokyo Denkai (Japan). The SCRF cavities were shipped to LANL for performance testing. This paper describes the experience gained during the manufacturing process at CERCA.

## 1 INTRODUCTION

Recently, a considerable interest emerged in the high-intensity proton linacs ranging in energy from 1 to 2 GeV because of the variety of possible applications, including the Accelerator Production of Tritium (APT), Accelerator Transmutation of Waste (ATW), and Spallation Neutron Source (SNS) [1]. These applications require powerful proton accelerator drivers capable of delivering to the target a proton beam of up to a hundred megawatts. The high beam power and continuous-wave (CW) regime of operation required by most applications make the accelerator design based on the SCRF technology desirable.

Elliptical SCRF cavities are successfully applied to accelerate electrons. However, for particles with $\beta<1$, the cavity shape poses some challenging problems, such as a higher likely of multipacting during the cavity operation. To answer these questions an Engineering Design and Demonstration (ED&D) program was initiated at LANL within the APT project. One goal of the ED&D program was to build, deliver, and test a complete cryomodule containing two medium-beta ($\beta = 0.64$), 700 MHz, 5-cell niobium SCRF cavities with helium vessels.

Supported by the U.S. Department of Energy under contract No. DE-AC04-AL89607.

Because the APT linac would require a large-scale application of superconducting technology, industrial participation in fabrication of SCRF cavities and other accelerator components is necessary. Burns and Roe Enterprises Inc. with General Atomics (BREI/GA) were selected by the U.S. Department of Energy (DOE) as the prime contractor for the APT project. Under a contract with DOE, BREI/GA coordinated the industrialization part of the ED&D program.

## 2 SCRF CAVITIES MANUFACTURING

### 2.1 Medium-beta SCRF cavity design requirements

Design of the $\beta = 0.64$, 700-MHz, 5-cell ED&D SCRF cavity underwent many iterations before the final version was released for fabrication [2]. Design work was supported by extensive structural and thermal analyses performed using the ABACUS® finite element code [3]. In addition, an experimental program to determine at 2 K the mechanical properties of Nb, Ti, and their mutual joints was initiated at Florida State University [4]. Information obtained from these measurements was essential to determining the structural requirements, i. e., maximum flaw size in the SCRF cavity and He vessel welds.

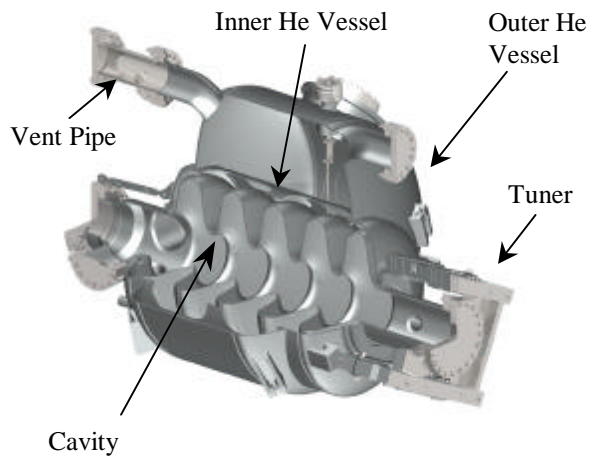

Figure 1. Cross section of the $\beta = 0.64$, 700-MHz, 5-cell ED&D SCRF cavity with helium vessels (computer rendering).

We present here the main features of the ED&D cavity design. The β = 0.64, 700-MHz, 5-cell ED&D SCRF cavity shown on Figure 1 is made from 4 mm thick solid Nb (RRR grade 250) sheets. There are two ports for power couplers (PC), two ports for higher-order mode (HOM) couplers, and one port for the RF pick-up. To improve RF power transmission to the cavity, the diameter of the PC beam tube was increased to 160 mm. Two tuner flexures are located on the beam tube cut-offs opposite the PC side. These are made from unalloyed grade 2 Ti.

The inner helium vessels (Figure 1) made from unalloyed grade 2 Ti are welded to the elliptical heads spun from unalloyed grade 2 Ti and to the beam tube through special Ti edge-welded bellows. The outer vessel has four access ports that may be welded shut after installation of cavity instrumentation.

Design requirements for manufacturing of ED&D cavities were specified in the Scope of Work (SOW) that was released to industry. These included the frequency of the fundamental accelerating mode (p-mode) to be 698.75 MHz at room temperature, RF energy density along the cavity axis (E-field flatness) to be within 10%, accelerating gradient $E_{acc}$ = 6.5 MV/m and the unloaded quality factor $Q_0 = 5\times10^9$ at $E_{acc}$ = 6.5MV/m. These last two were not a part of technical requirements for cavity acceptance, but were suggested design goals.

## 2.2 Fabrication of SCRF at CERCA

After winning a competitive bid process, CERCA was selected as the fabricator of four ED&D SCRF cavities. The cavities were manufactured in Romans, France. A detailed SOW that specified design requirements was submitted with the contract documents. CERCA provided manufacturing drawings and the manufacturing plan with detailed procedures for each manufacturing step, including an outline of the Quality Assurance plan based on the ISO 9000 approach. These were reviewed and accepted by the technical team from BREI/GA and LANL before cavity fabrication began.

Special attention was paid to provide sufficient quality controls at each step of the manufacturing process. All measurements and checks required by the manufacturing plan were performed and signed by the qualified QA technician and documented in the cavity traveller. When a nonconformance occurred, a detailed nonconformance report (NCR) was issued. The work stopped until the corrective action were proposed and accepted by BREI/GA and the LANL team. In addition, periodic visits to the CERCA manufacturing plant in Romans were made to provide oversight and ensure the quality of the manufactured parts.

CERCA manufacturing plan called for fabrication of half-cells by spinning. Two half-cells were spun to validate the process. Geometry of each cell was measured afterwards by a computerized coordinate measuring machine (CCMM) and found to be within specified tolerances of ±0.35 mm.

The cavity design requires single pass full-penetration cosmetic underbead electron-beam (EB) welds. The welding parameters were determined for each weld design and demonstrated on specimens of equal thickness. Since the equator weld is the most critical operation of the fabrication process, two spare half-cells were EB welded together at the equator to validate those weld parameters. The RRR of the parent, heat affected zone, and the welded material between high grades of niobium were measured at Oregon State University in Corvallis, Oregon [5] to ensure that the RRR did not degrade more than 10%. No RRR degradation was found within the accuracy of the measurement. [5]

Before starting the EB welding of the cavity parts, CERCA was required to provide specimens of high-to-low RRR Nb and low RRR Nb-to-Ti EB welded joints. These specimens were inspected to ensure that a complete penetration was achieved using these parameters. Once welding parameters were established and validated, CERCA was committed to maintain the weld parameters, and change them only after testing and acceptance by BREI/GA and LANL team.

Beam tube cut-offs and PC ports were made from Nb grade 250 material, while the HOM ports and RF pick-up tube were made from (reactor) RRR grade 40 material. Stainless steel CONFLAT® flanges were brazed to Nb tubes using a procedure developed by CERCA.

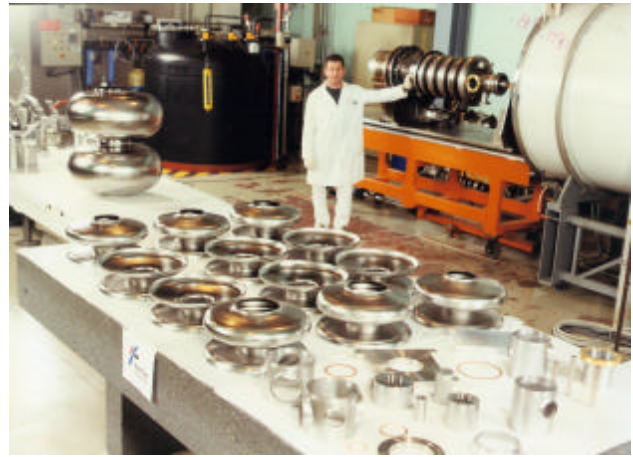

Figure 2. Fabrication of ED&D SCRF cavities at CERCA. In the foreground are "dumb-bells" formed by EB-welding of half-cells. In the background, a finished cavity is removed from the EB welder's chamber.

EB-welding of SCRF cavities followed the well-defined sequence described in the manufacturing plan. First, two half-cells were EB-welded (iris weld, performed from inside) to form "dumb-bells" (see Figure 2). Next, dumb-bells were EB-welded together (equator weld) to form an intermediate structure (see Figure 3). A single-pass, full-penetration EB-weld was performed from the outside. Finally, the beam tube cut-offs with end half-cells were EB-welded (equator weld) to form the five-cell cavity.

Before each EB-welding operation, parts were thoroughly cleaned (buffered chemical polishing, 1:1:1), rinsed in ultra-pure water, and dried under ultra-pure $N_2$. Adherence to this procedure was carefully checked and documented on the sign-off sheet in the traveller.

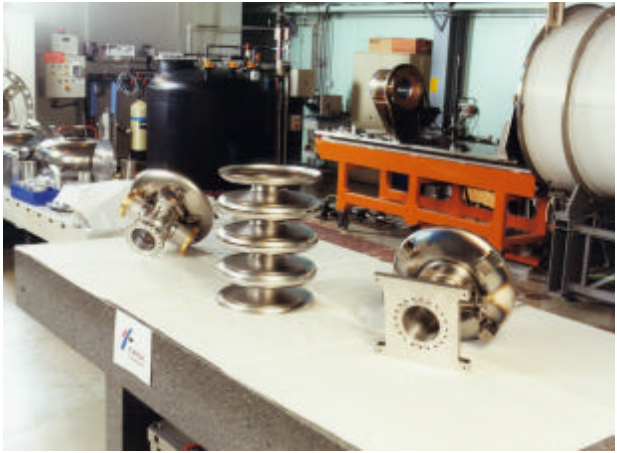

Figure 3. Beam tube cut-off and an intermediate part formed by EB-welding of dumb-bells prior to the final EB-welding.

Before shipment to LANL, all cavities underwent BCP (1:1:2) that removed 150 μm of Nb from the cavity inner surface. Cavities were filled with ultra-pure $N_2$ at 1.3 bar, sealed, and shipped to LANL in specially designed containers.

## 3 SCRF CAVITY PERFORMANCE

All four SCRF cavities manufactured by CERCA were cold-tested at 2 K to determine their performances. Cavities GERMAINE and SYLVIE were tested at Thomas Jefferson National Accelerator Facility [6], and ELEANORE and AYAKO were and are to be tested at LANL respectively [7]. Cavities were tuned to required frequency. RF energy density along the cavity axis (E-field flatness) was demonstrated to be better than 5%, thus exceeding design requirements. Prior to testing, cavities underwent a light, buffered chemical polishing that removed ~20 μm of Nb from the cavity's inner surface, followed by a high-pressure (90 bar) ultra-pure water rinsing.

Preliminary results for ELEANORE, GERMAINE, and SYLVIE are presented on Figure 4.

Maximum accelerating gradient $E_{acc}$ reached during tests was limited by the available RF power. As can be seen, performances of all three cavities exceeded the APT design goal. At the time this paper was written, testing of AYAKO was still underway.

After testing, the cavities were filled with ultra-pure nitrogen at a pressure of 1.3 bar, sealed, and shipped for the installation of the helium vessels.

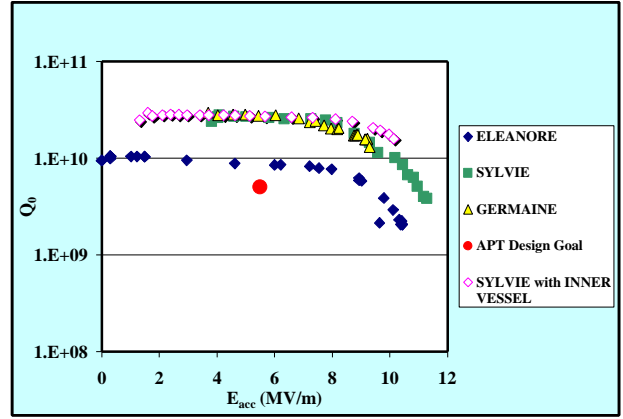

Figure 4. Performance of ED&D SCRF cavities manufactured by CERCA. ELEANORE and SYLVIE were built from Wah-Chang-supplied Nb and GERMAINE from Nb supplied by Heraeus AG.

## 4 CONCLUSION

During 1999, four SCRF cavities were successfully manufactured by CERCA for the APT project. During this process, a constant interaction between the customer and industrial partners resulted in high-quality product that exceeded the APT design goal. Preliminary results at 2 K show maximum accelerating gradient $E_{acc}$ in excess of 10 MV/m, the highest achieved to date in medium-beta multi-cell SCRF cavities. Industry experience injected into the manufacturing plan allowed simplification of the fabrication process and suggested cost-saving approaches.

## REFERENCES

[1] K.C.D. Chan, G.P. Lawrence, and J.D. Schneider, Development of RF Linac for High-Current Applications, NIM **B 139** (1998) 394-400.
[2] R.Gentzlinger et al.,"Fabrication of the APT Cavities," in Proceeding of the 18th Particle Accelerator Conference, New York, 1999.
[3] R.Mitchell et al., "Structural Analysis of the APT Superconducting Cavities", in IX Workshop on RF Superconductivity" Santa Fe, 1999.
[4] R.P.Walsh et al., "Low Temperature Tensile and Fracture Toughness Properties of SCRF cavity Structural Material", in IX Workshop on RF Superconductivity" Santa Fe, 1999.
[5] W.H.Warnes, private communication.
[6] J.Mammosser, private communication.
[7] T.Tajima, private communication.